\documentclass{llncs}[10pt]

\usepackage{fancyhdr} 
\usepackage{tikz}
\usetikzlibrary{arrows.meta}

\usepackage{amsmath}
\usepackage{amssymb}

\usepackage{lscape}
\usepackage{appendix}

\usepackage{mathtools}

\usepackage{comment}

\usepackage{setspace}
\usepackage{ntheorem}
\usepackage[ruled]{algorithm2e} 
\usepackage{graphicx}
\usepackage{enumerate}
\usepackage{enumitem}

\newtheorem{thm}{Theorem}    
\newtheorem{lem}{Lemma}  
\newtheorem{defi}{Definition}
\newtheorem*{prf}{Proof}

\usepackage{booktabs}

\usepackage[
	n,
	operators,
	advantage,
	sets,
	adversary,
	landau,
	probability,
	notions,	
	logic,
	ff,
	mm,
	primitives,
	events,
	complexity,
	asymptotics,
	keys]{cryptocode}
	
\usepackage{dashbox}
\usepackage{tikz}
\usepackage{enumerate}

\usepackage{enumitem}

\begin{document}

\title{Anonymous Blockchain-based System \\for Consortium\thanks{This work has been accepted by ACM TMIS \cite{wang2021anonymous}.}}

\author{
Qin Wang\inst{1,2}, Shiping Chen\inst{2}, Yang Xiang \inst{1}
}

\institute{
\textit{Swinburne University of Technology, Melbourne, Australia}
\\
\and
\textit{CSIRO Data61, Sydney, Australia} \\
\email{qinwang@swin.edu.au} 
}

\maketitle           
\begin{abstract}
Blockchain brings various advantages to online transactions. However, the total transparency of these transactions may leakage users' sensitive information. Requirements on both cooperation and anonymity for companies/organizations become necessary. In this paper, we propose a Multi-center Anonymous Blockchain-based (MAB) system, with joint management for the consortium and privacy protection for the participants. To achieve that, we formalize the syntax used by the MAB system and present a general construction based on a modular design. By applying cryptographic primitives to each module, we instantiate our scheme with anonymity and decentralization. Furthermore, we carry out a comprehensive formal analysis of the proposed solution.  The results demonstrate our constructed scheme is secure and efficient. 

\end{abstract}

\section{Introduction}
Blockchain technology, along with the widespread Bitcoin \cite{SN.2008}, has garnered insurmountable glories and attention from our society. More and more entrepreneurs and companies, including some world's biggest banks and technology firms, are investing massive amounts of resources, talent, and money to develop blockchain-based applications. Driven from the Bitcoin system, blockchain can be regarded as a distributed ledger with the advantages of irreversibility, pseudonymity, and unforgeability. The stability of the chain relies on miners, who use their electric power and computational hardware (i.e., consensus on PoW) or their digital stakes and public authorities (i.e., consensus on PoS, DPoS, PoA) to make the system proceed in consistency. However, the malicious behavior conducted by the miners and leaders, such as creating forks, may threaten the security of blockchain due to the centralization of resources. To avoid the issues, the concept of consortium blockchain has shown its superiority in many fields, including the joint management of organizations, the coin issuance of banks, and the censorship of governments. The multi-center technique can smoothly distribute centralized authorities into multiple entities, which brings relatively balanced powers across different entities. Also, the outcomes of the consortium blockchain will be beneficial to both academia and industry.

For academia, most of the studies in consortium blockchain focus on the implementation of joint control and supervision. For example, based on partially blind threshold signatures, Wu et al. \cite{Zhou2015} proposed a secure joint trading scheme on top of Bitcoin. The scheme allowed multiple participants to jointly manage one account, where no one could open it without supports from others. After that, by incorporating the secret-sharing method and attribute-based encryption scheme, Zhou et al. \cite{Zhou2016} proposed a Distributed Bitcoin Account Management (DBAM) framework to realize the hierarchically cooperated management scheme, especially suitable for enterprises and companies. Each account achieved fine-grained management among participants and prevented private keys from unexpected recovery. Crain et al. \cite{Crain2017} presented a targeting consortium blockchain based on the resilience optimal Byzantine consensus. This scheme described a simple and modular Byzantine consensus algorithm that mainly relied on randomization but not leaders and signatures. The algorithm simplified the problem of validating blocks, which makes it avoid the Sybil attacks.

For the industry, there are more and more companies and organizations working together to establish their own consortium blockchains. The Linux Foundation put forwards a PBFT based project called Hyperledger  \cite{androulaki2018hyperledger} to sever as the basic foundation of the blockchain. The idea was to create a cross-industry standard and an open-source development library so that business users could build their customized solutions, such as R3CEV's ledger, Corda \cite{r3corda}, etc. Corda fundamentally created an environment where each party could access the same data with the rules that all participants agree. Chinese companies also follow the above steps. The Chinaledger Alliance \cite{chinaledger} was sponsored by Wanxiang Blockchain Lab in 2015. They have created an open-source blockchain protocol and set standards across the industries to make sure the regulatory compliance can be satisfied. Moreover, the Post-Trade Distributed Ledger (PTDL) Group \cite{ptdl} improved the post-trade space which greatly benefits industrial members.

However, although numerous approaches of consortium blockchains are proposed and developed, they still confront the privacy-preserving problem across different parties in the consortium. More specifically, the plain contents (includes amounts, addresses, etc.) on the transactions can be freely accessed by the public, which may unconsciously leak some sensitive information of the members. For example, a malicious user can easily find the accounts with massive coins and trace the accounts to target the real entities by connecting them through some activities such as withdrawing coins by KYC, or exchanging coins offline. Such kinds of situations severely threaten the security of users' assets,  especially for the users like finical service providers in open banking \cite{xu2020ppm}.

To address the privacy problem, several approaches are proposed. For example, the confidential transaction, originally proposed by Adam Back \cite{Adam2013}, further developed by the Bitcoin Core Developer Gregory Maxwell \cite{Max2013,Max2015}, aims to enhance the privacy of transactions. Their idea was to embed the cryptographic commitment technique into the Bitcoin model. This idea was further modified by Shen et al. on Menero \cite{Shen2015}. In their implementation, Shen et al. employed the ring signature to confuse the participants from the input end, together with the Pedersen Commitment as the homomorphic tool to operate ciphertexts. Based on that, Ring-CT 2.0 \cite{CT2.0} was proposed to formalize the previous version and accelerate the computing efficiency via the cryptographic accumulator. Zerocoin and Zerocash \cite{zch2014} developed a method to confuse the sources of the transactions and change the mere pseudonymity into real anonymity, through the $mint$ and $pour$ step by using the non-interactive zero-knowledge proof (NIZKP). The sensitive information on the ledger was packaged into an envelope without leaking. Wang et al. \cite{WQ2017} proposed a privacy-preserving scheme to encrypt the original values from plaintexts to ciphertext through the homomorphic Paillier encryption system with the commitment proofs on its balance. The concept was designed as a compatible layer on top of the Bitcoin protocol. However, these solutions only fit the public blockchains, ignoring the scenarios of consortium. Therefore, considering both sides, the consequent question comes,

\textit{Could we establish a consortium blockchain for the group users while protecting the privacy of their sensitive data shown on transactions?}

Meeting such requirements is difficult. On the one side, due to the fact that a consortium blockchain grants several nodes with powerful authority, \textit{how} to balance the power of group members becomes the first concern. On the other side, blockchain is a public distributed ledger where all users can see and access the content of transactions. Protecting the privacy of sensitive data needs to seal the plaintexts into ciphertexts. This further brings questions on \textit{how} to ensure multiple ciphertexts can be correctly executed and \textit{how} to guarantee the value inside ciphertext is within the legal range. All these concerns should be fulfilled without any strong or extra assumptions. The proposed system should be compatible with various specifications and forms of applications. State-of-the-art systems can hardly well address such problems. For instance, Gai et al. \cite{gai2019privacy} proposed a consortium blockchain-oriented approach to solving the problem of privacy leakage of the group users that were located in nearby geographic positions. They mainly created noises (noise-based privacy-preserving approach) to hide the trading distribution trends for the system rather than screening individual privacy. However, the scheme only protects the system against a few types of attacks such as data mining-based attacks. This limits its application to other scenarios. Ra et al. \cite{ra2019anonymous} provided an anonymous protocol for privacy in a consortium blockchain. The protocol achieved privacy via transaction authentications by group members. This design is insufficient for universal applications because their solution relies on a strong assumption in which all the group members are trusted. This is unacceptable for real-world settings.  

\smallskip
\noindent\textbf{Contributions.} To solve the proposed issues, our system attempts to meet both requirements on decentralization (for consortium scenarios) and privacy (for sensitive information). Specifically, our motivation in this paper covers two aspects, 1) to establish a consortium chain for group members, and 2) to simultaneously protect their privacy of sensitive data recorded on-chain. To meet the first requirement, we need to make the involved parties or group members mutually combined through certain techniques (as shown in Joint Management Module). To meet the second requirement, we need to make the plaintext messages safely sealed through an encryption algorithm (as presented in Homomorphic Encryption Module). The question here is how to guarantee the encrypted values are correctly executed under the operations like addition and subtraction. Thus, the procedures of verification are necessary (as described in RangeVer Module and EqulityVer). Our scheme is designed and constructed on these decoupled functions. In this paper, we design a multi-center anonymous blockchain-based system (MAB system) to address the problems. Our construction is defined in formal modules and instantiated with several cryptographic primitives. The concrete construction is presented with security analyses and computation analyses. Here, we summarise the contributions of the paper as follows.

\begin{itemize}
  \item \textit{We propose a multi-center anonymous blockchain-based (MAB) system to solve the problems of decentralization and privacy.} The scheme is designed under the strictly formalized modules. Specifically, we have defined five modules as the basic elements to represent the functional blocks, covering the cooperation on the generation of trapdoors, the encryption on plain data, and the verification of ciphertexts. Consequently, by revoking the essential modules, we have presented the general construction of the MAB system under a strictly defined secure environment. Finally, we have specified each module with a practical algorithm to make the whole system flow through the blockchain-based system.

  \item \textit{We empower the MAB system with the advantages of decentralization, privacy, and efficiency.} Specifically, we have achieved the properties on 1) \textit{decentralization} by making participants jointly generate a system master key under the help of threshold key generation of RSA; 2) \textit{privacy} by realizing the concealment of the amounts of transactions through using homomorphic Paillier cryptosystem to hide the plaintexts, and the commitment proofs to keeping the balance and equality correct; 3) \textit{efficiency} by making the protocol easily and flexibly extended with auxiliary functions such as pre-processing, filter, etc.

  \item \textit{We provide formal security analyses on correctness, non-malleability, and balance.} Specifically, we have stated the rigorous security definitions and the corresponding proofs of the MAB system, where the results turn out to be secure under our requirements. Based on each customized module, the security guarantee of our proposed system is not only driven by the hard mathematical problems and assumptions, but also the robust blockchain system. According to the specific scheme, we proceed with the security definitions and proofs step by step, ensuring the properties on anonymity, non-malleability, and balance.
  

\end{itemize}

The rest of the paper is organized as follows. Section \ref{S::Cryptographic_Building_Blocks} provides the building blocks. Section \ref{S::System_Construction} presents the general construction of our MAB system. Section \ref{S::Security_Definitions} defines the security models. Section \ref{S::Instantiations} shows the concrete construction with the implementation skeleton. Section \ref{S::anaylse} provides security analyses and efficiency analyses. Section \ref{S::discussion} presents the related works with discussions. Finally, we conclude the paper in Section \ref{S::Conclusion}. Appendix A provides security proofs, Appendix B briefly introduces blockchain and Appendix C presents the system workflow.

\section{Cryptographic Building Blocks}
\label{S::Cryptographic_Building_Blocks}

In this section, we introduce the building blocks used in our construction.

\paragraph{RSA Keys in Threshold$^{\{A\}}$.}
The protocol aims to achieve the generation of a shared RSA key, that is, to make the consortium parties jointly generate a publicly known RSA modulus $N=pq$ without revealing its factorization. The encryption exponent is delivered to the public, and each party holds a share of the private exponent that enables threshold decryption. It is traditional to generate the parameter under a centralized party, however, the trust of authorities is always at worrisome risk. Numerous cryptographic protocols are requiring the RSA modulus $N=pq$ where none of the parties know its factorization. The original Fiat-Shamir protocol \cite{FS1987} uses modulus $N$ where no one knows its factorization, and the same does to \cite{Share1988}\cite{Share1996}. Threshold cryptography \cite{T-RSA2001} is an enlightened way to generate the shared RSA keys. The constructions that provide the $t-$out of$-k$ RSA threshold signature schemes can be found in \cite{Shared1991}\cite{Shared1998}.

\paragraph{Homomorphic Cryptosystem$^{\{B\}}$.}
Homomorphic encryption belongs to the public key cryptography system, and it is a cryptographic theory based on computational complexity and mathematical problems. Homomorphic operations include addition, multiplication, and other operations. Although full homomorphic encryption schemes \cite{fulHom2009}\cite{fulHom2010} realize arbitrary homomorphism of transactions, the extremely high cost of computation makes it out of reach. Two examples of Homomorphic Encryption schemes are additive ElGamal \cite{ElGam1985} and Paillier Encryption \cite{Paillier1999}, and both of them have been successfully applied in the design of e-voting schemes in the past. The additive encryption schemes are also employed by various state-of-the-art approaches on privacy-preserving protocols.

\paragraph{Commitment on Range Proof$^{\{C\}}$.}
Range proof is used to prove a committed value that lies in a specific interval. The scheme is based on the trapdoor commitment scheme which usually consists of two phases, 1) commit phase, and 2) reveal phase. The receiver sends a committed value in cyphertext, and then the sender can reveal the secret value for verification but the receiver knows nothing about the exact value. Range commitment confirms that the committed value lies in the rough interval without revealing the exact value. The proof was first proposed in \cite{Brickell1987}, and developed in \cite{Chan1998CFT,Boudot2000}. As shown in these references, CFT Proof \cite{Chan1998CFT} fulfilled the aim of proof, but it required an exceedingly large expansion rate from $[0,b]$ into $[-2^{120}b,2^{120}b]$ when compared to \cite{Boudot2000}. The method in \cite{Boudot2000} can limit the internal value to the range $[a,b]$ with negligible deviation, but the protocol is complicated. Wu et al. \cite{Zhou2015} has proposed a relatively efficient scheme with the proof content of $x\in[a; b]$ and the expansion rate was $\delta=1$.

\paragraph{Commitment on Equality Proof$^{\{D\}}$.}
The commitment to equality proof, sometimes called Indicative Commitment Proof, is a special scheme used to prove whether two committed values are equal or not. The traditional commitment specifies one committed value according to the trapdoor information, while the indicative commitment executes for two committed values. The unique property is that it allows two owners who hold trapdoor information to distinguish whether two committed values in the unreadable format are equal, without revealing commitments. The protocol is derived from equality proof of two discrete logarithms from \cite{Chaum1992,Chaum1998}, combined with knowledge proof of discrete logarithm `modulo $n$'. Moreover, the work \cite{Cam1999} provided a general form of an equality proof between two committed numbers in different moduli.

\begin{figure}[!hbt]
\centering
\includegraphics[width=\textwidth]{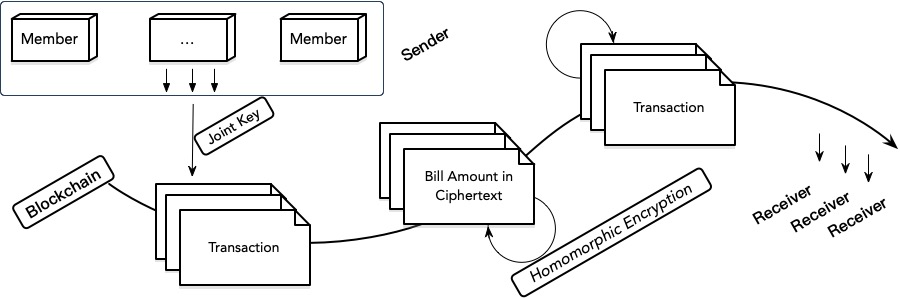}
\caption{Scheme Model: The figure briefly presents two major features of our scheme, jointly control and privacy protection. Group members obtain their separately shared key to co-generate the started parameter, and none of them can singly decrypt it. They use cokeygen parameters to encrypt messages transmitted on the chain, where the messages are confidentially protected from the public. After receiving the ciphertexts, the members will use their secret keys to decrypt them.}
\label{fig:1}
\end{figure}

\section{System Construction}
\label{S::System_Construction}
In this section, we briefly review basic cryptographic primitives used for our general construction and then present the formal definitions of each module towards their corresponding functional requirements. Based on rigorous definitions, we construct the complete general construction of the MAB system. The system is composed of five modules including \textit{Joint Management}, \textit{Homomorphic Cryptosystem}, \textit{Range Verification}, \textit{Equality Verification}, and \textit{Blockchain-based System}. We design these five modules according to key functions used in our scheme. Each of the modules separately represents a customized function to meet the requirements. Specifically, the homomorphic cryptosystem (module) realizes the function of making two encrypted values correctly executed with addition/subtraction operations. However, users, to their views, cannot know whether the output value is correct or not. Thus, another two verification modules are introduced. The range verification (module) solves the problems on whether an encrypted value is negative and the equality verification (module) ensures the sum of input values are consistent with the sum of operating output values. Similarly, the joint management (module) realizes the function where multiple participants are required to co-generate a secrete key for the subsequent decryption. Finally, the formalized security definitions are stated which contains anonymity, non-malleability, and balance. Here, we provide a high-level design of our system for simplicity as in Table.\ref{tab-highlevel}.

\begin{table}[!hbt]
 \caption{High-level Design}
  \centering
  \resizebox{0.9\linewidth}{!}{ 
    \begin{tabular}[t]{l|c|l|l}
    \toprule
     \textbf{Modules} &\textbf{Notation} & \textbf{Participated Entities} & \textbf{Functionality}\\  
    \midrule
    \emph{Joint Management} & $\Pi_1$  & Consortium Member & Secure bootstrap \\  \midrule
    \emph{Homomorphic Cryptosystem} &  $\Pi_2$  & Common User & Seal sensitive data \\  \midrule
    \emph{Range Verification} &  $\Pi_3$ & Common User & Prevent theft \\  \midrule
    \emph{Equality Verification} &   $\Pi_4$  & Common User  & Correctness/Integrity  \\ \midrule
    \emph{Blockchain} &  $\Pi_5$  &  Consortium Member & Secure platform \\ 
    \bottomrule
    \end{tabular}
    }
    \label{tab-highlevel}
\end{table}

\subsection{Construction on Modules}
\paragraph{Joint Management.}
 The module aims to corporately generate a shared private key. That is, the consortium parties jointly generate a publicly known RSA modulus $N=pq$ without revealing its factorization. We extend the vague model into a formal module $CoKeyGen$ based on the \textit{Building Block A}, to represent the process of \emph{Joint Management} in a real environment. Only when most of the consortium members (larger than the threshold) achieve an agreement, the factorization can be calculated to decrypt the ciphertext.

 \begin{defi}[CoKeyGen]
  The CoKeyGen module is composed of four probabilistic polynomial algorithms defined as \[\Pi_1=(rdmPara,coN,BipriTest,KeyGen)\] with the following syntax:
 \end{defi}
  \begin{itemize} [label=$\diamond$]
      \item rdmPara$(1^\lambda,k)$: \emph{When inputting a security parameter $\lambda$, each party $i$ secretly picks a secret $n-$bit integer $p_i$. The $p=p_1+...+p_k$ cannot be divisible via trial division by any prime in range. The second iteration for $q_i$ operates as the same.} 
      \item coN$(p_i, q_i)$: \emph{When inputting each party's parameter $p_i$and $q_i$, the algorithm outputs the public value of $N$ where $N=(p_1+...+p_k)(q_1+...+q_k)$. $N$ is not divisible by small primes in range.}  
      \item BipriTest$(N)$: \emph{When inputting a distributed computation by $k$ parties, the algorithm examines that $N$ is the product of two primes. If it fails, go back to step 1.} 
      \item KeyGen$(N,e)$: \emph{When inputting a public encryption exponent $e$, the $k$ parties jointly generate a secret decryption exponent $d$ as the key.}
  \end{itemize}

\paragraph{Homomorphic Encryption.}
  The module aims to encrypt metadata with the property of homomorphism for convenient calculation and operation. It can confidentially calculate the values in ciphertext while the obtains the correct results. We input the encrypted data and then process some operations to obtain an unreadable output. After the decryption, the results of the output data are consistent with the way they operate in plaintext.  According to the \textit{Building Block B}, we extended it into a formal module $HomoEnc$ to represent the process of \emph{Homomorphic Encryption}. 

 \begin{defi}[HomoEnc]
  The HomoEnc module is composed of four probabilistic polynomial algorithms defined as \[\Pi_2=(HKeyGen,HEnc,HOper,HDec)\] with the following syntax:
 \end{defi}
 
  \begin{itemize} [label=$\diamond$]
  \item  HKeyGen$(1^\lambda)$: \emph{When inputting a security parameter $\lambda$, the algorithm outputs public parameters $pms$ which can be implicitly a part of the input of the following algorithms.} 
  \item HEnc$(pk, m)$: \emph{when inputting a public key $pk_i$ and a massage $m$, the algorithm outputs an encrypted number $c_i=HEnc(pk_i, m)$.} 
  \item HOper$(c_i, c_j)$: \emph{When inputting several cyphertexts such as $c_i$ and $c_j$, the algorithm executes operations and outputs $c=c_i\otimes c_j$.}
  \item  HDec$(sk, c)$: \emph{When inputting a secret key $sk$ and the operated cyphertext $c$, the algorithm outputs $m=HDec(sk, c)$.}
   \end{itemize}

\medskip
The inherent homomorphism can be applied to various fields such as voting protocols, threshold cryptosystems, and secret sharing schemes. The scheme also possesses the properties of being self-reducible and self-blinding. Moreover, the cryptosystem is convincingly secure under the chosen-plaintext attack in the standard model such as the Paillier Cryptosystem, which is suitable for hiding the number of transactions on top of the blockchain-based system \cite{WQ2017}.

\paragraph{Range Verification.}
 The module aims to verify the hidden data lied in a specified interval $[a,b]$. The verification can correctly ensure the rough range without leaking any plain values. According to the \textit{Building Block C}, we extended it into a formal module $RangeVer$, to represent the process of \emph{Range Verification}. The verification improves privacy by hiding the metadata in transactions while ensuring the values are limited in the specific range.

\begin{defi}[RangeVer]
The RangeVer module is composed of four probabilistic polynomial algorithms defined as \[\Pi_3=(RKeyGen,RCom,RInact,RVer)\] with the following syntax:
\end{defi}

\begin{itemize} [label=$\diamond$]
\item RKeyGen$(1^\lambda)$: \emph{To input a security parameter $\lambda$, the algorithm outputs the public parameter $pk$, the random number $r$ and the trapdoor private key $sk$.}  
\item RCom$(pk, x)$: \emph{To input a public key $pk$ and a committed number $x$, the algorithm outputs a commitment $CM=RCom(pk,x)$.} 
\item RInact$(pk,CM,r)$: \emph{To input the parameter $pk,r$ and the commitment $CM$, the algorithm executes the interactive protocol and outputs the intermediate parameter $(w,s)$ on the reveal phase.} 
\item RVer$(w,s,CM,a,b,sk)$: \emph{To input a secret key $sk$, the commitment $CM$, and an expected interval $[a,b]$, the algorithm outputs $0/1$, where $1$ represents the committed number lies in the given interval, otherwise, it outputs $0$.}
\end{itemize}

\paragraph{Equality Verification.}
The module aims to verify the equality between the input sum and the output sum in transaction scripts. The verification will not leak any plain values while ensuring equality. According to the \textit{Building Block D}, we extended it into a formal module $BanlanceVer$ to represent the process of Equality Verification. The indicative features are reflected in the form of $0$ or $1$ rather than a concrete committed value. Only when the secret values inside commitments are equal, the algorithm outputs $1$, otherwise, outputs $0$.

\begin{defi}[EqualityVer]
The Equality module is composed of four probabilistic polynomial algorithms defined as \[\Pi_4=(EKeyGen,ECom,EInact,EVer)\] with the following syntax:
\end{defi}

\begin{itemize} [label=$\diamond$]
\item EKeyGen$(1^\lambda)$: \emph{To input a security parameter $\lambda$, the algorithm outputs the public parameter $pk$, the random number $r$ and the trapdoor private key $sk$.}  
\item ECom$(pk, x)$: \emph{To input the public keys $pk,pk'$ and the committed numbers $x,y$, the algorithm outputs unreadable commitments $C=BCom(pk,x)$ and $CM'=BCom(pk',y)$.} 
\item EInact$(pk,CM,CM',r)$: \emph{To input the parameter $pk,r$ and the commitment $CM,CM'$, the algorithm executes the interactive protocol and outputs the intermediate parameter $(w,s)$.}
\item EVer$(w,s,CM,CM',sk)$: \emph{To input a secret key $sk$, two commitments $CM$ and $CM'$, and parameters $w$ and $s$, the algorithm outputs $0$ or $1$, where $1$ represents two committed numbers inside commitments are equal, otherwise, the algorithm outputs $0$.}
\end{itemize}

\paragraph{Blockchain-based System.}
The blockchain-based trading system is inherently a distributed ledger with the properties of irreversibility, traceability, and persistence. It provides an environment for distributed applications. The system adopts the framework of Nakamoto \cite{SN.2008} to achieve credible records, distributed storage, and transparent transactions. It uses the token as the measurement unit to motivate the participants. The system is formed layer by layer through transactions, data blocks, and a unidirectional chain. The records on the generated blockchain are irreversible and persistent. We define it into a formal module $BChain$ in a simple way to capture the main structures of \emph{Blockchain-based System}.

\begin{defi}[BChain]
The BChain module represents the Blockchain-based System that has three layers stated as transactions, blocks, and a unidirectional chain, and each layer contains several sub-algorithms. The system is defined as
   \[  \begin{split}
         \Pi_5=& (Tx[Mint,Spend,Contract],Block[TxConfirm,Pour],\\
               &  Chain[Consensus,Recall,Expand])
       \end{split}
                     \] with the following syntax:
 \end{defi}

  \begin{description}
    \item[] Tx[Mint,Spend,Contract]: \emph{Tx} layer consists of three sub-algorithms defined as: 
    \begin{itemize} [label=$\diamond$]
     \item  Mint$(1^\lambda,TOut_{i-1},reward)$: \emph{To input a security parameter $\lambda$ for the key generation, the output value $TOut_{i-1}$ from last trading address and the $reward$ from miner if it is the first confirmed transaction, the algorithm outputs a public parameter $pk$, a trapdoor private key $sk$, and the total transaction value for verification if needed.} If it represents an enhanced privacy scheme with certain cryptographic primitives, the Mint of the coin needs to be integrated with the corresponding algorithms such as ring signature and zero-knowledge.   
     \item  Spend$(TIn_i,address)$: \emph{To input the results $Tin_i$, the algorithm sends coins to receivers according to its $pk$ or $address$.}
     \item  Contract$(\Theta,\mathcal{F})$: \emph{If the system is Turing-complete, it will execute a specific algorithm whenever the trigger conditions are satisfied. To input the trigger condition $\Theta$ and the algorithm $\mathcal{F}$, the initial states will be operated and transmitted according to its predefined rules.} The most mature Turing-complete system is Ethereum\cite{wood2014ethereum}, whose Distributed Apps(DAPPs) include ENS, Gnosis, Maker, etc.
     \end{itemize}

    \item[] Block[TxConfirm,Pour]: \emph{Block} layer consists of two sub-algorithms: 
    \begin{itemize} [label=$\diamond$]
      \item TxConfirm$(\mathbb{M},\mathcal{C},Tx)$: \emph{Entity $M\in\mathbb{M}$ verifies the existing transactions $Tx$ in the pool and generates a unique identification in hash-tree, which contains not only the amounts but also the possible contracts. There is only one winner $M_i$ who successfully builds up the block according to the consensus mechanism $\mathcal{C}$.} The confirmed block acts as an intermediate layer connect the transactions and chains. The entity $M\in\mathbb{M}$ represents the persons who successfully confirm blocks, including miners in POW, stakeholders in POS, candidates in DPOS, etc.
      \item  Pour$(\mathbb{T},\mathbb{A},\mathcal{S})$: \emph{According to the predefined strategies $\mathcal{S}$, the pour algorithm aims to mix/confuse the plain metadata in transactions $tx\in\mathbb{T}$ or a unique identification such as $address\in\mathbb{A}$.} Note that the pour process is designed especially for the cryptocurrency with strong privacy, and the component of mix represents different strategies, such as the mixed coins in Zcash\cite{kappos2018empirical}\cite{miers2013zerocoin} and the mixed addresses in RingCT\cite{noether2015ring}.
      \end{itemize}

    \item[] Chain[Consensus,Recall,Expand]: \emph{Chain} layer consists of three sub-algorithms: 
      \begin{itemize} [label=$\diamond$]
      \item Consensus$(\mathbb{M},\mathcal{C})$: \emph{Consensus $\mathcal{C}$ focuses on who can finally build the blocks and how to maintain the consistence. The inputs from entities  $\mathbb{M}$ 
      will be recorded on chain after an agreement according to its strategy $\mathcal{C}$.}  
      \item Recall$(\mathbb{F},\mathcal{C})$: \emph{`recall' represents the rollback of $forks\in \mathbb{F}$ according to the consensus mechanism $\mathcal{C}$.} Most forks converge into one main chain.
      \item Expand$(\Delta,\mathcal{E})$: \emph{`expand' contains the possible methods $\mathcal{E}$ with the aim to improve the scalability and performance. The methods includes attach side-chains to the main-chain such as the embedded sidechain and Lightning Network, or directly approve multiple parallel chains such as DAG-based blockchains.}
      \end{itemize}
  \end{description}

\subsection{Construction of MAB}

Based on the above-mentioned modules, we proceed to present the general construction of our Multi-center Anonymous Blockchain-based (MAB) system.

 \begin{defi}[MAB]
  The Multi-center Anonymous Blockchain-based System is composed of five modules as
  \[ 
         \Pi= MAB(CoKeyGen, HomoEnc, RangeVer, EqualityVer, BChain)
    \]
  which can also be denoted by the tuple as $ \Pi= (\Pi_1,\Pi_2,\Pi_3,\Pi_4,\Pi_5) $.
 \end{defi}

\smallskip
\fbox{
\parbox{\textwidth}{%
\begin{center}
   $\left.
    \begin{array}{ll}
     \emph{\underline{Transaction Flow:}}\\
       
     (p_m,q_m) \xleftarrow{\Pi_1} rdmPara(1^\lambda) \,\textrm{\small distribute parameters $p_i$ and $q_i$ for two iterations} \\

     (N,d_m) \xleftarrow{\Pi_1} CoN(p_m,q_m) \, \textrm{\small compute the public parameter $N$ } \\

     (0/1) \xleftarrow{\Pi_1}  BipriTest(N,\mathcal{T}) \,\textrm{\small biprimity test for integer $N$ under the strategy $\mathcal{T}$}\\

     (d) \xleftarrow{\Pi_1} KeyGen(N,e) \,\textrm{\small jointly generate the consortium private key }\\

     (pk_i) \xleftarrow{\Pi_5} Tx[Mint](d,1^\lambda,addr_i) \,\textrm{\small generate the public key for encryption}\\

     (c_i) \xleftarrow{\Pi_2} HEnc(pk_i,m_i) \,\textrm{\small homomorphically encrypt the metadata in transactions}\\

     (CM) \xleftarrow{\Pi_2} HOper(c_i,...,c_k) \,\textrm{\small homomorphic operation used for verifications}\\

     (w,s) \xleftarrow{\Pi_3} RInact(CM,pk,r) \,\textrm{\small execute the interactive protocol of range proof}\\

     (0/1) \xleftarrow{\Pi_3} RVer(w,s,[a,b],CM) \,\textrm{\small verify the validity in the specified range}\\

     (u,v) \xleftarrow{\Pi_4} EInact(CM,CM',pk,r') \,\textrm{\small execute the protocol of equality proof}\\

     (0/1) \xleftarrow{\Pi_4} EVer(u,v,CM,CM') \,\textrm{\small verify the equality between In-sum and Out-sum}\\

    (c_i^{addr_i}) \xleftarrow{\Pi_5} Tx[Spend](c_i,addr_i) \,  \textrm{\small transfer the coin from senders to receivers}\\

    (0/1) \xleftarrow{\Pi_5} Block[TxConfirm](M,Tx) \, \textrm{\small confirm Tx in blocks}\\

     (m_i) \xleftarrow{\Pi_2} HDec(c_i,sk_i)  \,\textrm{\small decrypt the ciphertext under the private key}\\

   \end{array}
   \right.$

\end{center}
 }%
}

The strictly defined modules could be substantiated by specific protocols according to the requirements. By calling the modules, our MAB system achieves properties and functions on the Joint Management for organizations, the Homomorphic Encryption for metadata, the Range Verification, and the Equality Verification for correct data. Here, we provide the complete workflow of MAB between the senders and the receivers by calling sub-algorithms and modules.

\section{Security Definitions}
\label{S::Security_Definitions}
Our MAB system, denoted as $\Pi$, aims to achieve the security properties of \textit{Anonymity}, \textit{Non-malleability} and \textit{Balance}, which are strictly defined as below. The adversary is represented as  $\mathcal{A}$ for short, and $Adv$ represents the advantage of an experiment. $pms,CM,m,r$ are the parameters, commitments, massages, and random numbers, respectively.

\paragraph{Anonymity.}
The property requires that the encrypted metadata shown on the transaction, such as the transferring value is completely hidden in a commitment $CM$, and the proof used in the protocol is at least computationally zero-knowledge. These two facts ensure that the adversary $\mathcal{A}$ has at most negligible advantage in guessing the plain information. 


 \begin{defi}[Anonymity]
  The MAB system $\Pi$ is anonymous-secure (ANY), if for all PPT adversaries $\mathcal{A}$ and sufficiently large $\lambda$, it holds that
 \[ \textrm{Adv}^{\textrm{ANY}}_{\Pi,\mathcal{A}}(\lambda)< \textrm{negl}(\lambda)  \]
 where $Adv^{ANY}_{\Pi,\mathcal{A}}(\lambda):=Pr[ANY(\Pi,\mathcal{A},\lambda)]$  is $\mathcal{A}$'s advantage in the experiment define as follows,
 \end{defi}


\begin{center}
   $\mathcal{A}(c)=b :\left[
    \begin{array}{ll}
    
     (CM,c)\gets HEnc/HOper/KeyGen(1^k);\\
     CM1\gets RVer(CM);\\
     CM2\gets EVer(CM);\\
     (c_0,c_1)\gets \mathcal{A}(CM1,CM2); \\
     b\gets \{0,1\};\\
     ^\star c\gets HDec(CM1,CM2,c_b) \\
     
   \end{array}
   \right].$
\end{center}

\paragraph{Non-malleability.}
The property requires that a malicious user cannot fake any transactions or others' identities after observing an honestly generated transaction. Namely, a commitment cannot be opened to two different values. Therefore, it is infeasible for attackers to produce a valid transaction or identity that shares the same information inside the ciphertext or the commitment. The non-malleability has already covered the linkability property, which aims to distinguish the difference between the participants.

 \begin{defi}[Non-malleability]
  The MAB system $\Pi$ is non-malleability(N-MAL), if for all PPT adversaries $\mathcal{A}$ and sufficiently large $\lambda$, it holds that
 \[ \textrm{Adv}^{\textrm{N-MAL}}_{\Pi,\mathcal{A}}(\lambda)< \textrm{negl}(\lambda)  \]
 where $Adv^{N-MAL}_{\Pi,\mathcal{A}}(\lambda):=Pr[N-MAL(\Pi,\mathcal{A},\lambda)]$  is $\mathcal{A}$'s advantage in the experiment define as follows. Note that, the left column is the preconditions of the experiment, and the right column shows the procedure of the experiment. The equation marked with $\star$ represents the condition used to decide whether $\mathcal{A}$ wins the experiment after the challenges,
 \end{defi}

\begin{center}
   $m_0\neq m_1 \land CM(m_0,r_0)=CM(m_1,r_1):\left[
    \begin{array}{ll}
    
     (pms)\gets KeyGen(1^k);\\
    ^\star (m_0,r_0,m_1,r_1)\gets \mathcal{A}(pms); \\

   \end{array}
   \right].$
\end{center}

\paragraph{Balance.}
The balance property consists of two parts, which requires that 1) any malicious user cannot spend negative value from others; technically, the committed numbers inside the ciphertexts should lie in a specific interval; and 2) each transaction can keep the balance and quality between inputs and outputs; technically, it is to prove that the quality of two unrelated committed values without revealing the commitments.

 \begin{defi}[Balance]
  The MAB system $\Pi$ is balance-secure(BAL), if for all PPT adversaries $\mathcal{A}$ and sufficiently large $\lambda$, it holds that
 \[ \textrm{Adv}^{\textrm{BAL}}_{\Pi,\mathcal{A}}(\lambda)< \textrm{negl}(\lambda)  \]
 where $Adv^{BAL}_{\Pi,\mathcal{A}}(\lambda):=Pr[BAL(\Pi,\mathcal{A},\lambda)]$  is $\mathcal{A}$'s advantage in the experiment define as follows,
 \end{defi}

\begin{center}
   $m_0, m_1 \in\mathbb{Z};r_0, r_1 \in\mathbb{Z}:\left[
    \begin{array}{ll}
    (pms)\gets KeyGen(1^k);\\
     c_0 \gets HEnc(pms,m_0,r_0),\\c_1 \gets HEnc(pms,m_1,r_1);\\
    ^\star 1\gets RVer(c_0,c_1);\\
    ^\star 1\gets EVer(c_0,c_1);\\

   \end{array}
   \right].$
\end{center}

\section{Instantiations}
\label{S::Instantiations}

\subsection{Concrete construction}
In this subsection, we present an instantiation of our MAB system under the formalized syntax. Our exemplified protocol extends the privacy-preserving protocol of Wang et al. \cite{WQ2017} with the feature of multi-centers. We specify each module with the customized algorithms for concrete construction. The steps of the modules are explained as follows.

\begin{description}
  \item[$\Pi_1:$] To jointly generate the trapdoor of RSA keys in transactions, technically, we employ the threshold scheme \cite{T-RSA2001} to divide a complete key into several shares and distribute them to multiple parties. The protocol publicly outputs an RSA modulus $N=pq$, but none of the parties know the factorization of $N$. Each participant holds a share of private exponents that enable the threshold decryption.

  \item[$\Pi_2:$] To hide the plain data in transactions, technically, we employ the homomorphic Paillier encryption scheme \cite{Paillier1999} to encrypt the plain amounts into ciphertexts. The scheme is based on the composite residuosity class problem, which is provably secure under the appropriate intractability assumptions in the standard model. The algorithm hides the original plain amounts, and only the receivers who own the private keys can decrypt the ciphertexts.

  \item[$\Pi_3:$] To ensure the positiveness of the encrypted values in transactions, technically, we employ the proof of committed numbers lied in specific intervals \cite{wu2004simple} to keep the value in range. The verification is essential to guarantee the security of individual accounts. This step prevents the deception inside the ciphertext from the attackers. 
      \[  \Pi_3=\textrm{PK}\{x,r|\;E=E(x,r)\bmod {n} \land x\in[a,b]\}  \]
      
  \item[$\Pi_4:$] To verify the equality between the input sums and the output sums, technically, we employ the proof of the equality of two committed numbers \cite{Boudot2000} to guarantee their consistency inside two blind commitments. We link it by ensuring the similarity of basic parameters, such as the generators of $g_d$ and $g_\beta$ in verification, and the random number $r_d$ and $h_\beta$ in encryption.
    \[
    \Pi_4= \textrm{PK}\{x,r_1,r_2| E=E_1(x,r_1)\bmod {n_1} \land  F=E_2(x_2,r_2)\bmod {n_2}\}  \\
   \]
\end{description}

$(p_m,q_m) \xleftarrow{\Pi_1} rdmPara(1^\lambda)$: Distribute parameters $p_m$ and $q_m$ for two iterations. Each party or group member $m$ selects an integer $p_m$ and keeps it secret. Then the $k$ parties compute $p=p_1+...+p_m+...+p_k$ where $p$ is indivisible by any primes less than some bound through a trial division test. If the step returns true, the parties execute the second iteration by $q=q_1+...+q_m+...+q_k$.

$(N,d_i) \xleftarrow{\Pi_1} CoN(p_m,q_m)$: The $k$ parties using a private distributed computation to compute the public parameter $N=(p=p_1+...+p_m+...+p_k)(q=q_1+...+q_m...+q_k)$. After that the parties further perform trail divisions to test whether $N$ is indivisible by any small primes in certain range.

$(0/1) \xleftarrow{\Pi_1}  BipriTest(N,\mathcal{T})$: The $k$ parties engage in a private distributed computation, via the biprimity test $\mathcal{T}$, to test the integer $N$ is the product of two primes. When this step returns true, then move into the next step. Note that the biprimality test is $k-1$ private.

$(d) \xleftarrow{\Pi_1} KeyGen(N,e)$: to input public encryption $e$, the parties jointly generate the consortium private key, and output a shared secret decryption exponent $d$. We can see that the RSA parameter $N$ is related to the setup of the system in the next step, and the secret exponent $d$ under the cooperated generation can be regarded as the trapdoor of the encryption system.

$(pk_i,pk_d) \xleftarrow{\Pi_5} Tx[Mint](1^\lambda,addr_i,N,p,q)$: For different receiver (distributed nodes) $i$, the system inputs the primes $p_i=p$, $q_i=q$, $n_i=N$, and generates the public key $pk_i=(n_i,g_i)$ for encryption, where $g_i\in \mathbb{Z}_{n_i^2}$ such that $g_i\equiv 1 \bmod n_i$, and secret key $sk_i=\lambda_i$ for decryption where $\lambda_i=\textrm{lcm}(p_i-1)(q_i-1)$. Simultaneously, the system generates another $pk_d=(n_d,g_d)$ as the parameter for the verification in \emph{dumb account} with no private key to spend money, where $n_d$ is a large safe composite number. The system then generates $V_\alpha(g_\alpha,h_\alpha)$ and $V_\beta(g_\beta,h_\beta)$, where $g_\alpha$ is an element of large order in $\mathbb{Z}_{n_\alpha}^*$ and $h_\alpha$ is an element of the group generated by $g_\alpha$ such that both the discrete logarithm of $g_\alpha$ in base $h_\alpha$ and the discrete logarithm of $h_\beta$ in base $g_\alpha$ are unknown to the sender. The same does to $g_\beta$ and $h_\beta$.

$(c_i,c_{id}) \xleftarrow{\Pi_2} HEnc(m_i,pk_i,pk_d)$:
The sender called Alice firstly computes the input-sum $m=m_{in}=\textrm{Dec}_{pk_{Alice}}(c_{in}\otimes 50reward)$(if Alice is miner). Employ the Paillier cryptosystem to homomorphically encrypt the plain amounts $m_1,m_2,...,m_i$ into ciphertexts $c_1,c_2,...,c_i$ under the different $pk_1,pk_2,...,pk_i$ from the receiver $i$. Simultaneously encrypt them under the system's $pk_d$ in parallel to the verification steps. Instead of being sent into the receivers' accounts, the encrypted bitcoins under the system $pk_d$ are sent into the \emph{dumb account}. Specifically, it goes as: $c_i=\textrm{Enc}_{pk_i}(m_i)=g_i^{m_i}r_i^{n_i}\bmod {n_i^2}$ in transaction layer and $c_{id}=\textrm{Enc}_{pk_d}(m_i)=g_d^{m_i}r_d^{n_d} \bmod {n_d^2}$ in verification layer. Note that the random number is $r_d=h_\beta$, which is an element of the group generated by $g_\beta\in \mathbb{Z}^*_{n_\beta}$. Clearly, system has encrypted the same amount $m_i$ in two layers with difference on public keys.

$(CM) \xleftarrow{\Pi_2} HOper(c_j,...,c_k)$: In the encryption process, Alice sends her coins to the receivers under $pk_i$, and sends the same amounts to the \emph{dumb account} under $pk_d$. The system executes homomorphic operation $CM=\prod c_{id}=HOper(c_id,...,c_kd)$ used for verification.

$(0/1)  \xleftarrow{\Pi_3} RInact,RVer(V_\alpha,V_\beta,CM,pk,m,m_i)$: Execute the interactive protocol of range proof. For simplicity, $g_d$, $h_d$, $n_d$ are written as  $g$, $h$, $n$, and the commitment made by Alice for each $m_i$: $E_{i0}(m_i,r)=g^{m_i}h_r \bmod n$, $E_{i1}$, $E_{i_2}$, $E_{i3}$, $F_i$, $V_i$ are written as $E_0(m_i,r)$, $E_1$, $E_2$, $E_3$, $F$, $V$ to represent a typical one of them. The algorithm goes: 1) Alice sends $(V,E_2,E_3,F)$ to Bob, where $E_1=g^{x-a}h^r=g^y h^r \bmod n$, $E_2=E_1^\alpha h^{r_1}\bmod n$, $E_3=E_2^\alpha h^{r_2}\bmod n$, $F=g^\omega h^{r_3}\bmod n$, $V=g^v/E_3=g^\omega h^{-r\alpha^2-r_1\alpha-r_2} \bmod n$. 2) Bob computes $E_1,U$, where $ E_1=E(x,r) /g^a=g^y h^r \bmod n$, $U=g^v/E_3=g^\omega h^{-r\alpha^2-r_1\alpha-r_2} \bmod n$. 3) Alice, Bob compute separately $\textrm{PK}_1,\textrm{PK}_2,\textrm{PK}_3$, where $\textrm{PK}_1\{ \alpha, r_1, r_2: E_2=E_1^\alpha h^{r_1}\bmod n \land E_3=E_2^\alpha h^{r_2}\bmod n\}$, $\textrm{PK}_2\{ \omega, r^*:\; F=g^\omega h^{r^3} \bmod n \land U=g^\omega h^{r^*} \bmod n\}$,$\textrm{PK}_3\{ \omega, r_3:\; F=g^\omega h^{r_3} \bmod n  \land   \omega \in [-2^{t+l+s+T}, 2^{t+l+s+T}]\} $. 4) Bob checks the correctness of PK$i(i=1,2,3)$  and that $v>2^{t+l+s+T}$, which convinces Bob that $x>a$ (we set $a=0$). 5) For each $m_i$, repeat step 1-4 to prove $m_i>0\;(i\in \{1,2,...,i\})$. Note that the steps will be iterated for $i$ times for every $m_i$ and if all succeed, the system returns 1, and then goes to the next.

$(0/1) \xleftarrow{\Pi_4} EInact,EVer(CM,CM',pk,r')$: Execute the interactive protocol of equality proof. To verify the output-sum $\sum m_i$ inside cooperated cipher $\prod c_i$ equals to input-sum $m$ in the commitment, we need to guarantee a)the secret $m=\sum m_i$ in $E$ and $F$ from Alice are identical, and b) the operated ciphertext $H=\prod c_{id}$ is equal to the committed number $F$. It proceeds as 1) Alice makes two committed number $E$ and $F$, where  $E=E_\alpha(m,r_\alpha)=g_\alpha^mh_\alpha^{r_\alpha}$, $F=E_\beta(\sum m_i,r_\beta)=g_\beta^mh_\beta^{r_\beta}$. 2) Alice computes $W_1,W_2$, where $W_1=g^\omega_1 h^{\eta_1}_1 \bmod {n_1}$, $W_2=g^\omega_2 h^{\eta_2}_2 \bmod {n_2}$, $\omega\in\{1,...,2^{i+t}b-1\}$, $\eta_1\in\{1,...,2^{l+t+s}n-1\}$, $ \eta_2\in\{1,...,2^{l+t+s}n-1\}$. 3) Alice sends to Bob with $(u,D,D_1,D_2)$, where $u=H(W_1||W_2)$, $D=\omega+ux$, $D_1=\eta_1+ur_1$, $D_2=\eta_2+ur_2$. 4) Bob checks whether $u=u'$, where $u'=H(g_1^Dh_1^{D_1}E^{-u}\bmod {n_1}||g_2^Dh_2^{D_2}F^{-u}\bmod {n_2})$. If successful, goes to next. 5) System in $dumb\;account$ computes $H$, where $H=\prod c_{id}=c_1'c'_2...c'_i=g_d^{m'_1+m'_2+...+m'_i}r_d^{n_d}=g_d^{\sum m'_i}r_d^{n_d}\bmod {n_d^2}$. 6) Compute $F$, where $F =g_\beta^{\sum m_i}h_\beta^{r_\beta} \bmod {n_\beta}=g_d^{\sum m_i}r_d^{n_d} \bmod {n_d^2}$, and parameters are specified as $r_d=h_\beta$, $r_\beta=n_d$, $n_\beta=n_d^2$, and $g_d=g_\beta$. 7) Check wether $H$ equals to $F$, where $ H=g_d^{\sum m'_i}r_d^{n_d} \bmod {n_d^2}$, $F=g_d^{\sum m_i}r_d^{n_d} \bmod {n_d^2}$. If yes, means $\sum m'_i=\sum m_i$, system returns $1$.

$(c_i^{addr_i}) \xleftarrow{\Pi_5} T_{Alice}[Spend](0/1,c_i,addr_i)$: The transaction, denoted as $T_{Alice}$, is broadcast to the P2P network to form a new block, and the contents on scripts are unrecognisable strings $c$ from the latest transaction and the related $addr_i$. The same pattern is suitable for all others. Note that before the transaction is spent, the \emph{dumb account} will be discarded automatically, since it exists only for the verification without private keys.

$(0/1) \xleftarrow{\Pi_5} Block[TxConfirm](M,T_{Alice})$: Miners $M$ verify the collected Tx, including $T_{Alice}$, into a confirmed blocks under the specific mechanism.

$(m_i) \xleftarrow{\Pi_2} HDec(c_i,sk_i)$: Receiver decrypt the ciphertext under the private key $sk_i$ as $m_i=\frac{L(c_i^{\lambda_i} \bmod {n_i^2})}{L(g_i^{\lambda_i} \bmod {n_i^2})} \bmod {n_i}$, where $L(x)=\frac{x-1}{n}$ and $x\in\mathcal{S}_n=\{u<n^2|x=1\bmod n\}$. The decrypted values are confidentially transferred from the sender to the receiver.

\subsection{Key Functions}
\begin{figure}[!htbp]
\centering
\includegraphics[width=0.9\linewidth]{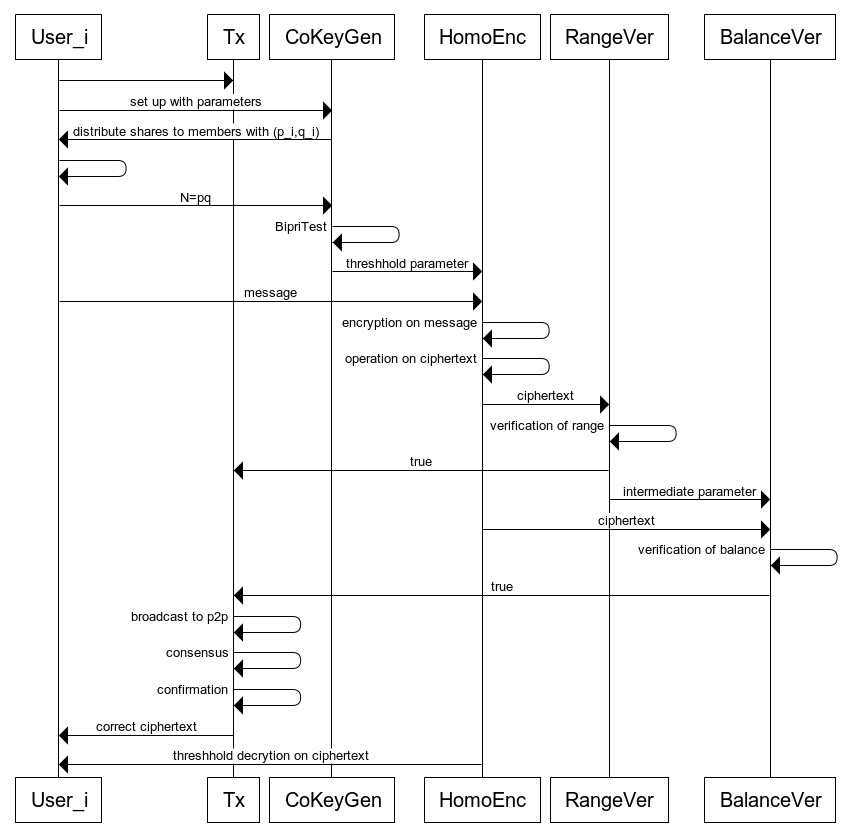}
\caption{The sequence diagram of scheme instantiation: Users jointly generate their threshold key-start parameters and then use them to encrypt the message on the blockchain. After receiving the encrypted message, group members have to reach an agreement to decrypt it. Note that we use $User_i$ to represent the group members.}
\label{fig:2}
\end{figure}

In this section, we emphasize two types of functions in our scheme. Here, we list their logic with the workflow in Fig.\ref{fig:2} at Appendix C.

\smallskip
\noindent\textbf{Function Design.}
We provide example codes, including \textit{CoKeyGen} and \textit{HomoEnc}, to show how we establish the functions. Our design presents the basic logic of the aforementioned modules.

\textit{Function of CoKeyGen}. 
This function attempts to achieve the generation of a shared RSA key. Consortium parties are required to corporately generate a publicly known RSA modulus $N=pq$ without revealing its factorization. Suppose there are $k$ parties ($k$ committee members), each party will generate their own private parameter $p_i$ and $q_i$. A systematic parameter is obtained by separately combining the parameter shares from committee members together, where any one cannot individually reverse the single share.  The BipriTest checks whether the generated systematic parameter is the product of two primes. If not passed, the algorithm returns. If passed, $k$ parties jointly generate a secrete decryption exponent as the systematic private key. Corporately generating the key by individual members indicates their mutual relationship as well as restrictions.

\begin{center}
\begin{algorithm}
\SetAlgoLined
\caption{CoKeyGen}
  \BlankLine
 \KwIn{Security parameter $\lambda$, public encryption exponent $e$}
 \KwOut{private key exponent $d$}
  \BlankLine

  \While{$1\leq i\leq k$}{
    $(p_i,q_i)$= rdmPara($1^\lambda, k$) \,\,\,\,\% \small generate key shares for each participants \\
  }
    N=coN($p_i,q_i$)\{       \,\,\,\,\% \small generate the systematic parameter \\  
    \quad $p=\Sigma p_i$\;
    \quad $q=\Sigma q_i$\;
    \quad $N=pq$ \;
    \}\;
    
  \While{$1\leq i\leq k$}{
  0/1=BipriTest(N)  \,\,\,\,\% \small check the conditions \\  
  \eIf{1}  
  {   return ``Passed"\;
   d=KeyGen(N,e)   \,\,\,\,\% \small check the systematic key exponent \\  
  }{return ``Failed"\;
  go to N=coN($p_i,q_i$)\;
  }
  }

\end{algorithm}
\end{center}

\textit{Function of HomoEnc}. 
This function attempts to encrypt the plaintext messages $m$ and decrypt the encrypted messages $c$. The encrypt metadata with the property of homomorphism for correct additive calculations of encrypted messages. The algorithm includes both the encryption procedure and the decryption procedure. For the encryption, we input the plain messages $m$ and the public key of users $pk$ and obtains the operated ciphertext $c$.  For the decryption, we input the encrypted messages $c$ and the public and private key pair of users $pk, sk$, and obtains the decrypted plaintext $m$. The results of the output messages are consistent with the way they operate in plaintext.

\begin{algorithm}
\SetAlgoLined
\caption{HomoEnc}
  \BlankLine
 \KwIn{public key $pk_i$, message $m$}
 \KwOut{-}
  \BlankLine
 
  $c$ = HEnc$(pk, m)$  \,\% \small encrypted the messages under the user's public key   \\   
  \While{True}{
    r = prime(round(math.log(pk.n, 2)))) \\
     \eIf{r > 0 and r < pk.n}{
            break }{return}
            }
    $x = pow(r, pk.n, pk.n_{square}) $ \;
    $c = (pow(pk.g, m, pk.n_{square}) * x) $ \;
    return $c$ \\
\}

~\\

 $m$= HDec$(sk, pk, c)$  \% \small decrypt the encrypted messages under the user's private key   \\
    $x = pow(cipher, sk.l, pk.n_sq) - 1$ \\
    $m = ((x // pk.n) * sk.m) $ \\ 
    return $m$\\
\}

\end{algorithm}


\smallskip
\noindent\textbf{Transaction Flow.} 
By calling the modules (functions) specified as above, we provide the work flow in Fig.\ref{fig:2} at Appendix C. For easy understanding, the structure of the system is presented in two layers including \textit{transaction layer} and \textit{verification layer}. Each layer will provide a specific function.

\section{Analysis of MAB}
\label{S::anaylse}

\subsection{Security Analysis}

In this section, we provide the security analyses of our protocol in detail.

\begin{thm}[Anonymity]
  Assuming that the discrete logarithm problem is hard in $\mathbb{G}_q$, the strong RSA assumption is hard, and the commitment proof is sound in the random oracle model, our proposed MAB protocol is anonymous.
\end{thm}

\begin{thm}[Non-malleability]
  Assuming that the discrete logarithm problem is hard in $\mathbb{G}_q$, the strong RSA assumption is hard, and the commitment proof is sound under random oracles, our proposed MAB protocol is non-malleable.
\end{thm}

\begin{thm}[Balance]
  Assuming that the discrete logarithm problem is hard in $\mathbb{G}_q$, the strong RSA assumption is hard, and the commitment proof is sound in the random oracle model, our proposed MAB protocol is balanced.
\end{thm}

As shown above, our scheme achieves the properties of anonymity, non-malleability, and balance. Detailed proofs are presented in Appendix A.

\subsection{Efficiency Analysis}

In this section, we provide the theoretical computation complexity of our scheme, including the sub-algorithms of Joint KeyGen, Encryption, Verification, Decryption, and Blockchain. We employ the $\tau_a,\tau_m,\tau_M,\tau_E,\tau_H$ to represent the unit times of the operations on addition, multiplication, modular multiplication, modular exponentiation, hash function operation, respectively. $\tau_{td}$ is for trial division test while $\tau_{bp}$ is for biprimality test. $\tau_{Tx}$ represents the P2P broadcasting time and I/O writing time, and $\tau_{Bl}$ is the block confirmation time by miners. The detailed evaluations are shown in Table \ref{tab-complexity}. Not that, these estimations are purely indicative, and are not derived from an actual implementation.

From the second column of Table \ref{tab-complexity}, we obtain that our scheme is theoretically efficient since each unit of $\tau$ is usually small in practice. The encryption and decryption take time at the same level. The verification costs most of the time since the operations are complex with the commitment proofs. The time of blockchain varies according to the underlying platform and its consensus.

\begin{table}[!hbt]
 \caption{Performance Evaluation}
  \centering
  \resizebox{0.7\linewidth}{!}{ 
    \begin{tabular}[t]{l|l}
    \toprule
    \textbf{Complexity} &\textbf{ Unit of Time } \\  
    \midrule
    Joint KeyGen & $\tau_{bp}+3\tau_{td}+2k\tau_a+\tau_m$    \\  \hline
    Encryption & $2i\tau_M+4i\tau_E$   \\  \hline
    Verification & $(2i+3)\tau_m+(7i+8)\tau_M+(12i+14)\tau_E+2\tau_H$  \\  \hline
    Decryption &  $2\tau_m+2i\tau_E$  \\ \hline
    Blockchain & $i\tau_{Tx}+\tau_{Bl}$  \\ 
    \bottomrule
    \end{tabular}
    }
    \label{tab-complexity}
\end{table}

\section{Related Work with Discussions}
\label{S::discussion}
\subsection{Types of Blockchain}
Generally, three types of blockchains are know to the public, namely \textit{private chain}, \textit{consortium chain} and \textit{public chain}. The rule to distinguish these three types of blockchain is the size of powerful members. A powerful member means the person who has the right to conduct key procedures of blockchain, such as packaging transactions, proposing blocks, and executing consensus. Private chain is the smallest blockchain in scale since only one person participates in the network. The person deploys the chain on his own and utilizes it for a brief simulation and experimental test, which is sufficiently flexible for further development. Any blockchain projects can be used as a private chain, like Ethereum \cite{wood2014ethereum},  Bitcoin \cite{SN.2008}, etc. Consortium chain and public chain have no restrictions on the size of members.  The key difference between the consortium chain and the public chain is whether they are permissioned. The consortium chain is permissioned, where two entities are involved, the committee node and the common node. A group of members, also called the committee, have the power of packaging blocks and executing consensus. Becoming the member requires meeting certain conditions. For example, schemes adopting PoS-based mechanisms make the users who hold the most stakes become the committee member. In contrast, the common node can only send and synchronize blocks from committee nodes. Famous consortium blockchain projects are Hyperledger \cite{androulaki2018hyperledger}, R3 Corda, and the Post-Trade Distributed Ledger (PTDL) Group (Webpage linkages can be found in the footnotes on Page 2). The public chain is permissionless, where all nodes can equally and freely join in or leave the network. This type of project maximumly improve the participation of the blockchain, since everyone can compete for rewards and profits. Typical projects are Bitcoin \cite{SN.2008} and Ethereum \cite{wood2014ethereum}.

\subsection{Consortium Blockchain}
Blockchain can be classified into three general types, private, consortium, and public blockchain \cite{bano2019sok}\cite{garay2020sok}. The private blockchain is deployed and maintained by a single organization, and it has weak centralization and low transaction costs. Public blockchain allows anyone to access the system and complete to generate the blocks and it is completely decentralized where the participants do not trust each other. Consortium Blockchain, as the balance between the private chain and public chain, specifies a group of members as the committee to maintain the chain. The consortium blockchain is suitable to establish decentralized applications with the features of multi-center regulation and high performance. Consortium blockchain is practically suitable for numerous fields, including art auction \cite{wang2019artchain}, financial service \cite{xu2020ppm}, etc.

To establish a consortium blockchain for group members, the priority is to 
establish a membership selection mechanism for the committee. The membership selection selects the nodes who can participate in the decision of consensus, to determine the blocks and the direction of the chain. Therefore, the (partially) permissioned consensus \cite{cachin2017blockchain}, with certain restrictions on node participation, is necessary for the system design. Pre-defined conditions in permissioned blockchains limit the size, behaviors, and power of the committee members. BFT-style protocols, as the most prevailing adopted consensus mechanisms in industries, serve for numerous companies and organizations like Linux Foundation, IBM's Hyperledger  \cite{androulaki2018hyperledger}, R3CEV's Corda, Chinaledger, etc. The protocols aim to address the Byzantine Problem in a distributed system. They solve the fault tolerance issues against the unpredictable behaviors of malicious nodes such as hardware errors, network congestion, and malicious attacks.  PBFT \cite{CastroL99} makes the BFT algorithm practical, and its variants are widely applied to blockchain systems. The solutions includes the dBFT of NEO \cite{wang2020security}, the BFT of Hotstuff \cite{yin2019hotstuff}, the PoS+PBFT of Tendermint \cite{amoussou2018correctness}, the vote-based BFT of Algorand \cite{gilad2017algorand}, etc.

\subsection{Privacy-Enhancing Blockchain}
The privacy of blockchain presents whether a blockchain can guarantee confidentiality for the data stored on it. For the blockchains purely targeted for trading and exchanging, the leakage of transaction information (amount, address, etc.) may exposure the privacy of users in real life. For the blockchain systems which employ the smart contract to support sophisticated logic, they confront additional risks in all layers. An adversary may attack the contracts by analyzing the vulnerabilities of deployed codes. Therefore, the privacy of sensitive data is urgently because the potential scope of blockchain is beyond merely virtual currencies. A primary example is to hide the bill amount in daily life \cite{wang2020blockchain}. 

There are several cryptographic tools used to enhance the privacy of the blockchain systems. 1) \textit{Crypto Commitment}: This method aims to hide the plaintexts by encrypted them without losing the feature of being operable. Therefore, it requires the commitment scheme to have homomorphic property. The original concept of the confidential transaction was proposed by Adam Back \cite{Adam2013}, further developed by Gregory Maxwell \cite{Max2013,Max2015}.  Their idea was to embed the cryptographic commitment technique into the Bitcoin model. Wang et al. \cite{WQ2017} proposed a scheme by using the homomorphic Paillier encryption system with the commitment proofs on its balance. 2) \textit{Ring Signature}: It is a type of digital signature used to hide the identity in the group. This idea was firstly adopted by Menero \cite{Shen2015}. The scheme confuses the participants at the input end, together with the Pedersen Commitment as the homomorphic tool to operate ciphertexts.  Later, the improved Ring-CT \cite{CT2.0} was proposed to enhance the protocols with higher efficiency and better privacy. 3) \textit{Zero-Knowledge Proof (ZKP)}:  ZKP represents a series of protocols with the aim to prove the correctness of specified value to someone without exactly revealing it. Zerocoin/Zerocash \cite{zch2014} was the first blockchains system to adopt the ZKP technique. It proposed a method to confuse the sources of the transactions, which changes the mere pseudonymity into real anonymity. The sensitive information on the ledger was hidden through the steps of $mint$ and $pour$.

\section{Conclusion}
\label{S::Conclusion}
In this paper, we propose a Multi-center Anonymous Blockchain-based system under the strictly defined security environment, which provides the properties on  1) improving the privacy of metadata presented on the ledgers; 2) realizing the joint management in the consortiums. To achieve the system, firstly, we provide the general construction by formalizing five modules based on cryptographic primitives to reconstruct the functional blocks of our scheme, which achieves the cooperation on the generation of trapdoors, the encryption on plain data, and the verification on ciphertexts. Secondly, we customize each module by employing relatively matured algorithms to build the concrete construction. The modules employ the techniques on the threshold RSA, the Paillier cryptosystem, and the commitment proof. Finally, we present the rigorous security definitions and the corresponding proofs of our scheme. The results demonstrates that the proposed scheme is strictly secure and practically efficient. 

\smallskip
\noindent\textbf{Acknowledgement.} Sincere appreciations to my previous supervisor Prof. Qianhong Wu (Beihang University, China) for his ever discussion with me on the idea. Part of the construction has followed my previous solution in \cite{WQ2017}. Also, this work has been partially completed during my Master's study at Beihang University.

\bibliographystyle{unsrt}
\bibliography{bib}

\section*{Appendix A: Proof for Theorems}

\subsection{Proof for Theorems 1}

\begin{prf}[Anonymity]
The proof proceeds via the following games. We set a serious $\Game_{\textrm{i}}$ where the adversary $\mathcal{A}$ interacts with a challenger $\mathcal{C}$, as in the ANY experiment. More precisely, we describe a sequence of games $(\Game_{\textrm{real}},\Game_{1},\Game_{2},\Game_{3},\Game_{\textrm{4}})$ to show that  $Adv^{ANY}_{\Pi,\mathcal{A}}(\lambda)$ is negligibly different from that in the real experiment.  We define $\Game_{\textrm{real}}$ to be the original experiment. We also define $adv^{ver}$ to be the advantage in encryption's experiment, $adv^{prf}$ to be the advantage in distinguishing the pseudo-random function $prf$ from a random one, and $adv^{ver}$ to be the advantage towards the hiding property of commitments.

\textbf{Game $\Game_{1}$:} The game $\Game_{1}$ modifies $\Game_{real}$ by replacing the ciphertexts with the encryptions on random stings. We modify $\Game_{real}$ so that $\mathcal{C}$ simulates each encryption process. More specifically, once $\mathcal{A}$ outputs ciphertext $c$, there is $c'=\varepsilon(pk,r)$ to be generated where $r$ is the uniformly selected messages from the $m$ in plaintext space. By the Lemma 1, we get \[ |adv^{\Game_{1}}-adv^{\Game_{real}}|\leq negl(\lambda).  \]

\textbf{Game $\Game_{2}$:} The game $\Game_{2}$ modifies $\Game_{1}$ by calling the simulator to replace the $pms$ and all signatures of knowledge in commitment proofs. We modify $\Game_{1}$ so that $\mathcal{C}$ simulates each commitment proofs. More precisely, instead of invoking $KeyGen(1^{\lambda},RVer,EVer)$, $\mathcal{C}$ invokes $Sim=(1^{\lambda},RVer,EVer)$. For each invocation of the verification algorithm, $\mathcal{C}$ computes $\pi_{ver}\gets Sim(trap,m)$ without using any witnesses. Since the commitment proof is perfect zero knowledge shown in \cite{Fujisaki1997,Zhou2015,Chan1998CFT}, the simulated proof $\pi_{ver}$ in is identical to it computed in the real experiment. Therefore, we get \[ |adv^{\Game_{2}}-adv^{\Game_{1}}|\leq negl(\lambda).  \]

\textbf{Game $\Game_{3}$:}  The game $\Game_{3}$ modifies $\Game_{2}$ by confusing its addresses selection. Intuitively, the game differs from the previous one during the challenge phase when the output address is chosen uniformly at random. More specifically, in this game we set $h_i=h_{i-1}^\alpha$, and the distribution parameter $r$ is uniformly at random. Thus, based on the hidden property with homomorphism, we get that \[ adv^{\Game_{3}}=adv^{\Game_{2}}. \]

\textbf{Game $\Game_{4}$:} We describe the game $\Game_{\textrm{4}}$ as follows. Set the $b\in\{0,1\}$ at random, $\mathcal{C}$ samples $pms\gets KeyGen$. As in the ANY experiment, $\mathcal{C}$ sends $pms$ to $\mathcal{A}$, and then initialises two oracles $\mathcal{O}^{\textrm{MAB}}_1$ and $\mathcal{O}^{\textrm{MAB}}_2$. $\mathcal{C}$ provides $\mathcal{A}$ two responses separately in $\mathcal{O}^{\textrm{MAB}}_b$ and $\mathcal{O}^{\textrm{MAB}}_{1-b}$ for each step running in the experiment. Then $\mathcal{A}$ sends to $\mathcal{C}$ a message $(Q,Q')$ which contains two kinds of queries in the same type.  According to the query type $q$, $\mathcal{C}$ acts in slight difference:
\begin{equation*}
\begin{array}{ll}
  q_{Mint}: Q=Q'=Mint(add_{ac})=a \\
    \quad\textrm{employ a random string $\nu$ to replace the $a_{pk}$ in address, }\\
    \quad\textrm{the same does to $Q'$}  \\
  q_{Enc}: Q=(Enc,pms,add_{ac},m),\\
    \quad\quad\quad Q'=(Enc,pms,add'_{ac},m) \\
    \quad \textrm{employ $c_{Enc}=(\nu||pms||m)$ instead of $c_{Enc}=(a||pms||m)$,}\\
    \quad\textrm{the same does to $Q'$} \\
  q_{RVer}: Q=(RVer,pms,add_{ac},m), \\
    \quad\quad\quad Q'=(RVer,pms,add'_{ac},m) \\
    \quad \textrm{employ $C_R=(\nu||pms||m)$ instead of $C_R=(a||pms||m)$, }\\
    \quad\textrm{the same does to $Q'$}\\
  q_{EVer}: Q=(EVer,pms,add_{ac},m), \\
    \quad\quad\quad  Q'=(EVer,pms,add'_{ac},m) \\
    \quad \textrm{employ $C_E=(\nu||pms||m)$ instead of $C_E=(a||pms||m)$, }\\
    \quad\textrm{the same does to $Q'$} \\
  \end{array}
  \end{equation*}
It is clear that the response to $\mathcal{A}$ is independently of $b\in \{0,1\}$. Therefore, $\mathcal{A}$'s advantage in $\Game_{\textrm{4}}$ is 0 since the view is no difference from $b$ or $b'=1-b$ where $Pr[b=b']=1/2$. So we get  \[ |adv^{\Game_{4}}-adv^{\Game_{3}}|\leq negl(\lambda).  \]

\begin{lem}
Under the CCRA assumption, Game $\Game_{1}$ is computationally indistinguishable from Game $\Game_{real}$. More precisely, for all PPT adversaries $\mathcal{A}$ and sufficiently large $\lambda$, it holds that
\[   |adv^{\Game_{1}}-adv^{\Game_{real}}|\leq negl(\lambda)   \]
where $adv^{Enc}$ is the maximum of $\mathcal{A}$'s advantage in the encryption scheme, and $q_e$ represents encryption queries.
\end{lem}

\begin{prf}
Suppose that there exists an adversary $\mathcal{A}$ that can distinguish $\Game_{1}$ and $\Game_{real}$ with non-negligible probability. We can then design an efficient distinguisher $\mathcal{D}$ to solve the CCRA problem \cite{Paillier1999}. CCRA represents the Computational Composite Residuosity Assumption which states that: given $c\in \mathbb{Z}_{n^2}^*$ and $n=pq$, there is no polynomial-time algorithm to successfully compute $m$ in $c=g^mr^n \mod n$ where $m\in \mathbb{Z}_n$ and $r\in \mathbb{Z}_{n^2}^*$. The reduction from this to standard CCRA assumption draws the conclusion of the indistinguishability.

Without loss of genarity, we assume that $\mathcal{A}$ makes at most $q_k$ and $q_e$ queries to HKeyGen and HEnc respectively. For some $j\in\{1,...,q_{k}\}$ when $\mathcal{A}$ makes the $j$-th query of the HKeyGen, query the challenger to obtain the public keys $(pk_0,pk_1)$ in response to $\mathcal{A}$. At the time $\mathcal{A}$ issues the $q_e$ query resulting in the ciphertext $c_i$ under $pk_0$, while query the challenger to get the ciphertext $c_i'=\textrm{HEnc}(pk_{\overline{b}},m)$ under $pk_{\overline{b}}$, where $\overline{b}$ is the bit chosen by challenger. $\mathcal{A}$ outputs $b'$ as a guess in the experiment. When $\overline{b}=0$, $\mathcal{A}$'s view of the interaction is identical to that of $\Game{real}$, while when $\overline{b}=1$, $\mathcal{A}$'s view represents the intermediate simulator that the key is substituted. Eventually, $\mathcal{D}$ outputs 1 if $b'=b$.

We assume that $adv^{Enc}$ is the maximum of $\mathcal{A}$'s advantage in the encryption scheme, $Evt_1$ represents the first event mention above while $Evt_2$ towards the second, then we can induce that
  \begin{align*}
    adv^{Enc}=& |adv^{\Game_{1}}-adv^{\Game_{real}}|   \\
    =& Pr[\mathcal{D}(c) = 1|Evt_1 \land Evt_2]-
     Pr[\mathcal{D}(g^mr^n) =1 |Evt_1 \land Evt_2]  \\
    =&\frac{Pr[\mathcal{D}(c) = 1] - Pr[\mathcal{D}(g^mr^n) = 1]}{Pr[Evt_1 \land Evt_2]} \\
    =&2q_e\cdot adv^{CCRA}_{\mathcal{D}}  \\
    \leq& negl(\lambda)
 \end{align*}
 \end{prf}

Hence, under the CCRA assumption and the securities of building blocks in each module, our protocol is anonymous as shown above.
\end{prf}

\subsection{Proof for Theorems 2}

\begin{prf}[Non-malleability]
Set $\mathbb{T}$ be the set of transactions generated by $\mathcal{O}^{\textrm{MAB}}$ in response to queries, recall that $\mathcal{A}$ wins the N-MAL experiment where there exists $tx\in \mathbb{T}$ satisfies: on input $tx\neq tx'$, then output $Verify(pms,tx',pk)=1$. To achieve the goal, a malicious attacker needs to 1) break the encryption system by solving the hard problem of strong RSA after the KeyGen step, denoted as Event$_{RSA}$; 2) break the commitments by solving the hard problem of DL in Verification step, denoted as Event$_{DL}$. To simulate the attack, we employ a random RSA and a DL instance under the simulated games separately, such that if there exists an adversary that can slander honest spending, then the designed protocol can be reduced to the mathematical problem. We suppose there is an efficient adversary $\mathcal{A}$ that can break the non-malleability of our protocol with non-negligible probability $\epsilon$. We can use $\mathcal{A}$ as a subroutine to design an efficient algorithm $\mathcal{A}^*$ to solve the hard problem in $\mathbb{G}_q$.

\textbf{Event$_{RSA}$:} The strong RSA problem states as follow: given an RSA modulus $n$ and a random element $\nu \in \mathbb{Z}^*_n$, it is hard to find $e>1$ and $u$ such that $z=u^e$. Now, to reduce the Event$_{RSA}$ to the hard mathematic problem above, we simulate the following interaction. Given the random strong RSA instance $(N,u)$, the algorithm $\mathcal{A}^*$ selects $(g,h)$ and the co-generated parameters $(p,q)$ in $\Pi_1$ as the KeyGen, where the parameters can be set as $(N,u,p,q,g,h)$. It generates the $(c_1,...,c_k)$ in $\Pi_2$ and $\Pi_3$. To reduce the event of dishonesty, $\mathcal{A}^*$ produces valid output $(\pi, CM)$ and $c'=c_j^* \notin \mathbb{C}$ by the extraction from $\pi$, note that $j^*\in[q_a]$ is randomly picked by $\mathcal{A}^*$. $\mathcal{A}$ extracts $\omega$ from $\pi$ and employs the resulting value to compute the solution of strong RSA instance. Here, we conclude the advantage probability which states as follow
 \begin{align*}
    adv^{Event_{RSA}}= & q_e\cdot adv^{RSA}   \\
    \leq & q_e\cdot Pr[adv^{\mathcal{A}(\overline{abort})}] \cdot Pr[ j^*\in S
         \cap S^*|\overline{abort} ]\cdot Pr[\overline{abort}] \\
    =    & q_e\cdot Pr[adv^{\mathcal{A}} \land j^*\in S \cap S^*|\overline{abort} ]\cdot Pr[\overline{abort}] \\
    \leq & q_e\cdot Pr[\mathcal{A}^*(m,(CM'^d)^e) = m] \\
    \leq & negl(\lambda)
 \end{align*}

\textbf{Event$_{DL}$:} Given the random DL instance $(h_0,h_1)$ where $h_1=h_0^{\alpha}$, the algorithm $\mathcal{A}^*$ inputs $pms$ by running the KeyGen. Then, $\mathcal{A}^*$ randomly picks $j^*\in[q_a]$ and randomly selects $x_i\in \mathbb{Z}_q$, so that the address can be driven as $pk_i=[h_0^{x_i}|i \neq j^*] $ while $pk_i=[h_1^{x_i}|i = j^* ]$. Note that $sk_{j^*}=\alpha\cdot x_{j^*}$ inside the equation in the revoking process. After the setting, $\mathcal{A}^*$ simulates $\mathcal{O}^{\textrm{MAB}}$ in the experiment. At the end of the algorithm, $\mathcal{A}$ outputs $(S,tx,\pi)$ and $(S^*,tx^*,\pi^*)$. If $\mathcal{A}$ succeeds in the experiment with the requirements 1) $tx^*\neq tx$ and $S \cap \S^*=\emptyset$ , and 2) $Verify(pms,tx^*,pk)=1$, it means $\mathcal{A}$ can solve the DL problem behind the algorithm. Technically, check whether $j^*\in S \cap S^*$, outputs a random guess if it not; otherwise, use the trapdoor to extract the valid witness $(S^*,tx^*,\pi^*)$, where it contains secret key $sk_{j^*}$ that $pk_{j^*}=h_0^*=h_0^{\alpha\cdot x_j^*}$, so that $\alpha=sk_{j^*}/x_{j^*}$ as the DL solution of $h_1$ to $h_0$.

Eventually, we suppose that $\mathcal{A}^*$ succeeds to solve the DL problem, and set $abort$ as the event that $\mathcal{A}^*$ aborts while $\overline{abort}$ as the complementary event. If $abort$ does not happen, the simulated experiment is the same the the real from the $\mathcal{A}$'s view. If $j^*\in S \cap S^*$, $\mathcal{A}^*$ can successfully solve the discrete logarithm, which obtains $\alpha$ from $h_1=h_0^{\alpha}$. From the above, we reduce the Event$_{DL}$ to hard problem and conclude its advantage probability as follows:
  \begin{align*}
    adv^{Event_{DL}}= & q_a\cdot adv^{DL}   \\
    \leq & q_a\cdot Pr[adv^{\mathcal{A}(\overline{abort})}] \cdot
         Pr[ j^*\in S \cap S^*|\overline{abort} ]\cdot Pr[\overline{abort}] \\
    =    & q_a\cdot Pr[adv^{\mathcal{A}} \land j^*\in S \cap S^*|\overline{abort} ]\cdot Pr[\overline{abort}] \\
    \leq & q_a\cdot Pr[adv^{\mathcal{A}} \land j^*\in S \cap S^* ]  \\
    \leq & q_a\cdot Pr[\mathcal{A}^*(h_0,h_1^\alpha) = \alpha] \\
    \leq & negl(\lambda)
 \end{align*}

Hence, under the RSA and DL assumption, the protocol is non-malleable in the random oracle.
\end{prf}

\subsection{Proof for Theorems 3}
\begin{prf}[Balance]
According to the requirements, the balance property contains both range verification and equality verification. Intuitively, for each transaction, $\mathcal{A}$ did not spend money more than the amount that $\mathcal{A}$ has, and the amount from the receivers should match the senders. Formally, we define $\epsilon:=Adv^{BAL}_{\Pi,\mathcal{A}}(\lambda)$, and our goal is to show that $\epsilon$ is negligible in $\lambda$. The transactions induced by the BAL experiment are balanced with all but non-negligible probability. Suppose, by the way of contradiction, $\mathcal{A}$ can induce, with non-negligible probability, a transaction ledger that is not balanced. According to aforementioned requirements, we show how to reach a contradiction via our commitment proofs and we explain Verify$_{Range}$ and Verify$_{Equality}$ in sequence.

\textbf{Verify$_{Range}$:}

The receiver Bob was acknowledged by PK$_1$ that
  \begin{align*}
    E_1&=g^y h^r \bmod n \\
    E_2&=E_1^\alpha h^{r_1}=g^{\alpha y}h^{\alpha r+r_1} \bmod n \\
    E_3&=E_2^\alpha h^{r_2}=g^{\alpha^2 y}h^{\alpha^2 r+\alpha r_1+r_2} \bmod n
  \end{align*}
 and was also acknowledged by PK$_2$ that
  \begin{align*}
    F&=g^\omega h^{r_3}\bmod n  \\
    V&=g^v/E_3=g^\omega h^{-r\alpha^2-r_1\alpha-r_2} \bmod n
  \end{align*}
 so that
  \begin{align*}
   g^v&= VE_3   \\
      &= g^\omega h^{-r\alpha^2-r_1\alpha-r_2}g^{\alpha^2 y}h^{\alpha^2 r+\alpha r_1+r_2}  \\
      &= g^{\alpha^2 y+\omega}h^0    \\
      &= g^{\alpha^2 y+\omega}   \bmod n
  \end{align*}
 and from the equations we get
  \[ v=\alpha^2 y+\omega  \mod{\varphi(n) } \]
where $\varphi(n)$ is a Totient function. Here, the sender Alice knows neither the factorization of $n$ nor the function $\varphi(n)$. From the strong-RSA assumption, there only exists $ v=\alpha^2 y+\omega $.  Since Bob has verified that $v>2^{t+l+s+T}$ and convinces from PK$_3$ that
 \[ \omega\in [-2^{t+l+s+T},2^{t+l+s+T}], \]
so we get
 \[y>0 .\]
Otherwise, if $y<0$, we obtain
 \[ v=\alpha^2 y+\omega \leq \omega \leq 2^{t+l+s+T}\]
which \textbf{contradicts} to $v>2^{t+l+s+T}$ in the \emph{RangeVer} step.

From the above, we can prove $y=m_i-a>0$, and $m_i>a$, where $a=0$. The correctness analysis can be applied to each $m_i$ for $i\in \{1,2,...,i\}$. Therefore, $v_{addr\to\mathcal{A}}\geqslant 0$ for each sender, and after that we go into the next:

\textbf{Verify$_{Equality}$:} From the verification step \cite{Boudot2000}, we can see
 \[ u=H(W_\alpha||W_\beta)=(g_\alpha^\omega h_\alpha^{\eta_\alpha}\bmod n_\alpha|| g_\beta^\omega h_\beta^{\eta_\beta}\bmod n_\beta) . \]
 and if two committed numbers $E$ and $F$ are equal, we have:
 \begin{align*}
  u'=& H(g_\alpha^Dh_\alpha^{D_\alpha}E^{-u}\bmod {n_\alpha}||g_\beta^Dh_\beta^{D_\beta}F^{-u}\bmod {n_\beta})    \\
    =& H(g_\alpha^{um+\omega}h_\alpha^{ur_\alpha+\eta_\alpha}({g_\alpha^mh_\alpha^{r_\alpha}})^{-u}\bmod {n_\alpha}||
      g_\beta^{um+\omega}h_\beta^{ur_\beta+\eta_\beta}({g_\beta^mh_\beta^{r_\beta}})^{-u}\bmod {n_\beta}) \\
    =& H(g_\alpha^{um+\omega-um}h_\alpha^{ur_\alpha+\eta_\alpha-ur_\alpha}\bmod {n_\alpha}||
      g_\beta^{um+\omega-um}h_\beta^{ur_\beta+\eta_\beta-ur_\beta}\bmod {n_\beta})  \\
    =& (g_\alpha^\omega h_\alpha^{\eta_\alpha}\bmod n_\alpha|| g_\beta^\omega h_\beta^{\eta_\beta}\bmod n_\beta)  \\
    =& u  .
 \end{align*}
and then, we focus on the operated ciphertext $H$ and the committed number $F$ from Alice:
\begin{align*}
              H &=\prod c'_i  
                =g_d^{\sum m'_i}r_d^{n_d} \bmod {n_d^2}    \\
              F &=g_\beta^{\sum m_i}h_\beta^{r_\beta} \bmod {n_\beta}
                =g_d^{\sum m_i}r_d^{n_d} \bmod {n_d^2}
\end{align*}
where  $r_d=h_\beta$ , $r_\beta=n_d$ are randomly selected , and $n_\beta=n_d^2$, and $g_d=g_\beta$ are generated above. If $H$ equals to $F$ where
\[  H=g_d^{\sum m'_i}r_d^{n_d} \bmod {n_d^2}=g_d^{\sum m_i}r_d^{n_d} \bmod {n_d^2}=F   \]
we can clearly conclude that:
\[  \sum m'_i=\sum m_i=m \]
which means that the input-sum equals to the output-sum, and the verification turns to be equal. Otherwise, if the accounts that are either from the senders' or the receivers' will be attacked by the $\mathcal{A}$, the commitments $H$ and $F$ will change in accumulation by the exponential operations, which leads to an inside significant difference as the \textbf{contradiction}. Therefore, $v_{reward}+v_{unspend}+v_{addr\to\mathcal{A}}=v_{\mathcal{A}\to addr}+v_{self}$.

As shown above, the adversary $\mathcal{A}$ can neither steal coins from others by inputting a negative value nor attack the systems by changing the amounts in dishonesty. Therefore, the proposed MAB system both passes the verification of range and equality, which ensures the balanced security of the system.
\end{prf}
\section*{Appendix B: Skeleton of Blockchain}

\begin{figure}[!htbp]
\includegraphics[width=\textwidth]{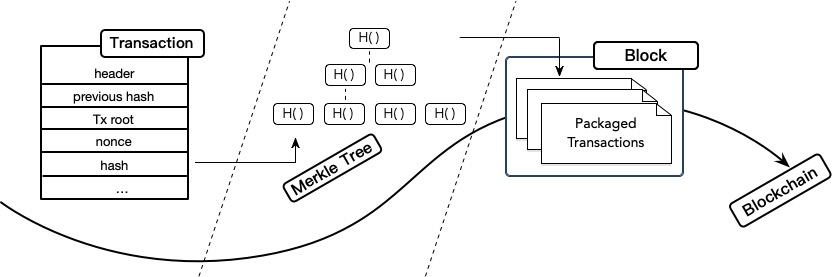}
\caption{Skeleton of the Blockchain-based System}
\label{fig:3}
\end{figure}

Blockchain can be regarded as a distributed ledger to record the activities in the form of transactions, then these transactions are organized into a hierarchical structure as a block, and last the blocks are arranged in an irreversibly ordered sequence. Each block is guarded by cryptography techniques to provide a strong guarantee of security. The way to compete for the rights of packaging transactions relies on a series of rules called t the consensus mechanism. New blocks can only be committed into the main chain when the consistency of decisions reaches. Therefore, blockchain ensures that once a block is committed, the transaction cannot be tampered with and compromised, and consequently the integrity and correctness of the data recorded on transactions are guaranteed. Fig.\ref{fig:3} provides a skeleton of the blockchain-based system. This is the most prevailing structure adopted by current blockchain systems \cite{SN.2008}\cite{wood2014ethereum}\cite{wang2020security}\cite{yin2019hotstuff}\cite{bano2019sok}. Other structures such as DAG-based blockchains systems refer to \cite{wang2020sok}.

\section*{Appendix C: System Workflow}

\end{document}